\documentclass[11pt]{article}
\usepackage{graphicx}
\usepackage{amssymb}
\usepackage{amscd}
\usepackage{mathrsfs}
\usepackage{longtable,lscape}
\usepackage{amsthm}
\usepackage{amsfonts}
\usepackage{amsmath}
\usepackage{bbm}
\usepackage{float}
\usepackage{url}
\usepackage{hyperref}
\usepackage[margin=0.5cm,font=footnotesize]{caption}
\usepackage{lineno}
\usepackage[round]{natbib}
\usepackage{appendix}
\usepackage{subfig}

\begin{document}
\title{\bf The Separability Problem in Quantum Mechanics: Insights from Research on Axiomatics \\ and Human Language}
\author{Diederik Aerts$^*$, Jonito Aerts Arg\"uelles$^*$, Lester Beltran$^*$, \\ Massimiliano Sassoli de Bianchi\footnote{Center Leo Apostel for Interdisciplinary Studies, 
        Vrije Universiteit Brussel (VUB), Pleinlaan 2,
         1050 Brussels, Belgium; email addresses: diraerts@vub.be,jonitoarguelles@gmail.com,lestercc21@yahoo.com,autoricerca@gmail.com} 
        $\,$ and $\,$  Sandro Sozzo\footnote{Department of Humanities and Cultural Heritage (DIUM) and Centre CQSCS, University of Udine, Vicolo Florio 2/b, 33100 Udine, Italy; email address: sandro.sozzo@uniud.it}              }

\date{}

\maketitle
\begin{abstract}
\noindent 
Einstein's article on the EPR paradox is the most cited of his works, but not many know that it was not fully representative of the way he thought about the incompleteness of the quantum formalism. Indeed, his main worry was not Heisenberg's uncertainty principle, which he accepted, but the experimental  {\it non-separability} of spatially separate systems. The same problem was also recognized, years later, by one of us, as part of an axiomatic analysis of the quantum formalism, which revealed an unexpected structural limitation of the quantum formalism in Hilbert space, preventing the description of separate  systems. As we will explain, this limitation does not manifest at the level of the states, but of the projectors describing the properties, in the sense that there are not enough properties in the formalism to describe separate systems. The question remains whether  {\it separability} is a possibility at the fundamental level and if a formalism should integrate it into its mathematical structure, as a possibility. To aid our intuition, we offer a reflection based on a powerful analogy between physical systems and human conceptual entities, as the question of separability also arises for the latter. 
\end{abstract}

\section{Introduction\label{intro}}

The ``intellectual duel'' between Einstein and Bohr is perhaps the most emblematic in the history of modern physics  \citep{einstein1946,bohr1949}. It has been described by many as the confrontation between a strenuously conservative Einstein, struggling to preserve the values of classical physics, and a more modern Bohr,  ready instead to abandon all nostalgia for the `old classical world' and willing to fully open to the mystery and strangeness of the `new quantum world'; see for example \citep{pais1982,despagnat1983,fine86}. 

In the imagination of most physicists, Einstein sought in every way to preserve the narrative of classical physics by battling Heisenberg's uncertainty principle. All his thought experiments ({\it gedankenexperiments}) seemed in fact to suggest, at least if read superficially, that the main enemy was precisely such principle, which in the famous EPR article \citep{epr1935} was believed guilty of contradicting a reality criterion of great generality, apparently indicating that position and momentum were perfectly well defined physical quantities, for every single microscopic entity. Hence, quantum mechanics could only be incomplete. 

However, in the often overly distorted description of the intellectual contraposition  between these two giants, it is easy to lose sight of the nuances. But it is in these nuances that it is possible to grasp the true nature of their important dispute and the value of Einstein's concerns and critiques of quantum mechanics, always profound and full of insights, whose significance has often not been adequately understood by his contemporaries and, we might add, probably by Einstein himself,  as he could not always match his insights with precise mathematical results.

Speaking of the 1935 EPR article \citep{epr1935}, it is worth underlining that it was written by Podolsky and not by Einstein, who was particularly unhappy with the way Podoslky presented the issue. Indeed, that same year, in a letter to Schr\"odinger, he wrote: ``For reasons of language this [paper] was written by Podolsky after several discussions. Still, it did not come out as well as I had originally wanted; rather, the essential thing was, so to speak, smothered by the formalism [gelehrsamkeit].'' The reason for Einstein's ``repudiation'' of his own article is contained in those nuances mentioned above, which are usually not found in the typical account of his critique of the quantum formalism. 

One of these nuances, perhaps the most important, was described in great detail by the philosopher Don Howard, an expert on the work of Einstein and Bohr \citep{howard1985,howard1990}. According to Howard, what most commentators on Einstein's work failed to understand is that the main issue he discussed with Bohr was that of \emph{non-separability}. In fact, Einstein wasn't particularly worried about Heisenberg's uncertainty principle and never designed his many {\it gedankenexperiments} to try to refute it. He had reservations, to be sure, about its ontological status, but he considered the principle a secondary concern compared to the issue of the `non-separability of quantum interactions'. As reported in \citet{howard1990,howard2005}, this can be clearly inferred from a letter that Paul Ehrenfest wrote to Bohr in July 1931, after visiting Einstein in Berlin, mentioning a comment from Einstein: ``He [Einstein] said to me that, for a very long time already, he absolutely no longer doubted the uncertainty relations, and that he thus, e.g., had \emph{by no means} invented the `weighable light-flash box' [...] `contra uncertainty relation', but for a totally different purpose.'' 

This `totally different purpose' was that of showing that the quantum mechanical narrative was in denial of \emph{separability}, that is, the requirement that two spatially separate systems possess their own separate reality, hence well-defined separate states. Therefore, in his paradigmatic debate with Bohr, Einstein should not be distorted as a nostalgic old man with a strong idiosyncrasy towards indeterminism, but as a scientist who had profound insights into the quantum formalism and its relation to the physical world, and the fact that a fundamental incompleteness was possibly hidden in quantum mathematics, in relation to separability. This means that his 1935 paper should be re-read having in mind the \emph{separation principle} and not the uncertainty principle. In other words, quoting from \citet{howard1985}, ``by Einstein's own account, the central issue in his dispute with Bohr concerns the separability principle. Bohr defends the completeness thesis by repudiating the separability principle; Einstein defends separability by denying completeness.''

The importance Einstein attaches to the {\it separability principle},  i.e., to the `independent existence of spatially distant physical systems', can be traced back to his conception of what should be the fundamental ingredients of a valid field theory, including his theory of general relativity. In fact, according to him, field theories are by definition separable theories. In general relativity, for example, the metric tensor is well defined at every point in the space-time manifold. Quoting \citet{Einstein1948}: ``Field theory has carried out this [separability] principle to the extreme, in that it localizes within infinitely small (four-dimensional) space-elements the elementary things existing independently of one another that it takes as basic, as well as the elementary laws it postulates for them.''

Another reason why Einstein's attention was focused on separability from the early 20th century onwards was his contributions in the formulation of the new (Bose-Einstein) quantum statistics\footnote{The Bose--Einstein statistics is a procedure for counting the states of systems composed of indistinguishable entities with integer spin.} and the fact it implied a lack of statistical independence of the microstates of identical quantum entities \citep{howard1990}. This lack of statistical independence, a consequence of \emph{quantum correlations},\footnote{Quantum mechanics predicts that joint measurements on composite systems in entangled states reveal correlations between observed values. The pattern exhibited by these correlations, however, seems to conflict with classical intuitions, particularly the idea that there may be a common cause in the past at the origin of the correlations. These seem instead to indicate that a composite system in an entangled state behaves more as an indivisible whole than as two separate subsystems, and that the observed correlations are created and not discovered by the act of observation.} led Einstein to conjecture the existence of a mysterious force at the origin of the phenomenon, whose nature, however, remained to be clarified and remains an open question to this day. In \citet{ijtp2023,philtransa2023,aertsetal2024paper1}, we proposed to understand this mysterious force, which brings together identical \emph{bosons},\footnote{In physics, bosons are entities obeying the Bose--Einstein statistics. Some elementary bosons act as force carriers, like photons for the electromagnetic force and gluons for the strong nuclear force.} as the physical counterpart of the ``meaning force'' active in a text that tells a story, and the associated mechanism of \emph{contextual updating} carried by meaning. This powerful analogy, between `quantum behavior' and `cognitive behavior', will be further explored in the final part of this article, to shed light on the question of the existence, or non-existence, of separate physical systems and the necessity of having theories that, from a structural point of view, can accomodate to describe such situations. 

It is important to emphasize that the separability of two systems is not equivalent to requiring that there is no interaction between them. The principle of separability, as understood by Einstein, arises at a more fundamental level, as a principle of individuation of physical systems, which in some way makes it possible to establish whether we are in the presence of two systems (distinguishable precisely because they are separable) or of a single system. And in the latter case it is no longer possible to speak of an interaction between the two systems, because, precisely, we are no longer dealing with two systems. Also, Einstein's principle of separation is to be understood as being the result of a `spatial separation', in the sense that two systems would become independent of each other, i.e., describable independently of each other regarding their states, if they become sufficiently spatially separate (typically, by a space-like interval). 

Here we can emphasize a classical prejudice cultivated by Einstein, that spatial separation always implies experimental separation, i.e. an independence of the outcomes of measurements performed on two separate systems. In other words, Einstein separability is to be understood in the sense of `an independent existence of spatially distant entities'. And let us mention a second classical prejudice also cultivated by Einstein, that `to exist' means necessarily `to be localized in space'. 

Coming back to the EPR paper, Einstein was not happy with Podolsky's formulation precisely because it failed to make explicit this central assumption of separability and, as Howard rightly points out \citep{howard1985}, this left the door open to the kind of criticism that Bohr made, for example pwhen in his response to the EPR article he wrote the following cryptic sentence \citep{Bohr1935}: ``[...] there is essentially the question of an influence on the very conditions which define the possible types of predictions regarding the future behavior of the system. Since these conditions constitute an inherent element of the description of any phenomenon to which the term `physical reality' can be properly attached, we see that the argumentation of the mentioned authors does not justify their conclusion that quantum-mechanical description is essentially incomplete.'' What Bohr refers to here is the fact that the system considered, with the associated experimental setting, is not to be considered a composite system formed by two separate sub-systems. So, again, the central element of the dispute between Einstein and Bohr is precisely separability. 

As Howard emphasizes \citep{howard1985}: ``The EPR argument seeks to prove quantum mechanics incomplete by proving the existence of elements of physical reality having no counterpart in the theory; it is thus an attempt at a direct proof of incompleteness. By contrast, Einstein's own argument does not require the identification of specific elements of physical reality not mirrored in the theory. Einstein's argument seeks, instead, to exhibit a contradiction between the completeness assumption and the consequences of the `separation principle', which makes it an indirect proof of incompleteness.''  More precisely, Einstein considers quantum mechanics `incomplete' because it assigns different vector states to describe a same `state of affair' of a physical system, as deduced from the `separation principle' and the application of his reality criterion, plus a reasoning based on counterfactual conditionals.

Now, as much as Einstein had the right intuition in believing that the incompleteness of quantum mechanics was to blame for quantum non-separability (as we will have a chance to point out very clearly in this article), i.e., the existence of quantum correlations despite the separation of systems (according to his prejudice that spatial distance was sufficient to guarantee it), nevertheless Einstein could not understand at his time at what level exactly the problem was situated, in the quantum formalism. Similarly, the reasoning reported in the EPR article, which Einstein considered misleading because it masked the central importance of the hypothesis of non-separability (focusing instead on the simultaneous reality of incompatible physical quantities, such as position and momentum), which he believed was necessary for the independent existence of different systems, remained a reasoning {\it ex absurdum}, thus unable to specify which of the assumptions underlying the reasoning was the one responsible for the contradiction. 

There was also an additional difficulty. With regard to the description of states, there seems to be in the quantum formalism a clear way of describing a composite system consisting of two separate sub-systems. One simply has to describe the state of the composite system as a \emph{product state}, and product states give rise to the expected factorization of the probabilities. Hence, Einstein's concern about quantum non-separability seems at first sight unfounded, being it possible to describe separate quantum entities by putting them in a product state. Einstein certainly received this objection in his time. We find evidence of it, for example, in a letter from Pauli to Heisenberg where he complained about the damage the EPR article could do. In it, he asked Heisenberg to write a response to the article, specifying, among other things, the possibility of describing two independent sub-systems in terms of a decomposition of the eigenfunctions into a product \citep{Pauli1985}. 

The authors of this article also often received the objection that the quantum non-separability problem could not be such, since, precisely, it is possible to use product states to describe  separate  systems. However, this objection is very weak. First of all, because one should be aware that quantum evolution brings a product state describing two apparently separate interacting systems into a non-product (entangled) state just in an instant. In other words, quantum evolution in the presence of an interaction does not conserve product states.

But as we will explain in detail in this article, the situation is much more subtle than that. In accordance with Einstein's insights, it is correct to say that quantum mechanics fails to describe separate quantum entities, but the failure does not appear at the level of the states.The tensor product provides enough states to describe separate entities, using product states, but what is missing in the formalism is not states, but properties, in the sense that there are properties that are peculiar to composite systems formed by separate sub-systems, that can be made perfectly explicit, which cannot be described as closed subspaces (projection operators) of the tensor product Hilbert space \citep{aerts1981,aerts1982}. In other words, there are elements of reality that standard quantum mechanics is unable to describe and this means that there is indeed a serious structural problem in the quantum formalism, and in some way Einstein was able to perceive it correctly, although not to express it fully, as the mathematics to do so was not available in his day. 

The development of this new mathematics required the pioneering work of Birkhoff and von Neumann \citep{BirkhoffvonNeumann1936}, which gave rise to a new field of inquiry focused on being able to obtain an axiomatic foundation of quantum mechanics, possibly realistic and operational. This work found its greatest development in the 1970s and 1980s with the work of Mackey, Jauch, Piron, Foulis, Randall and Ludwig \citep{Mackey1963,Jauch1968,Piron1976,FoulisRandall1978,Ludwig1983}. One of the authors, a student of Constantin Piron, has been very actively involved in this quantum axiomatics research over the past century, so much so that the Geneva school of quantum mechanics has become known -- at least among insiders -- as the \emph{Geneva-Brussels school of quantum mechanics}  \citep{aerts1982,aerts1986,aerts1999a,aertsdurt1994,aertsetal1997a,aertsetal1999b}.\footnote{One should not think of the Geneva-Brussels School as a physical institution where formal education or research takes place, but rather as an intellectual tradition in which scholars pursue a common approach in exploring the foundations of physical theories and, more recently, of human cognition. The name is primarily due to the pioneering contributions of Ernst St\"{u}ckelberg, Jospeh Maria Jauch and Constantin Piron, and their collaborators at the University of Geneva, and the subsequent developments achieved by Diederik Aerts and his collaborators at the Vrije Universiteit Brussel.}

It was within the scope of this school's work that it was possible to frame standard quantum mechanics within more general mathematical structures, able to jointly describe quantum and classical systems, understanding them as limit situations of more general `in between quantum and classical' intermediary systems. This made it possible to appreciate that physical systems have properties that can be classical, quantum or intermediate, regardless of whether they are microscopic or macroscopic, or immersed in a hot or cold environment. In other words, the term ``quantum'' refers first and foremost to a form of organization present in certain experimental settings and revealed through the performance of appropriately designed experiments. 

In this more general framework, it was also possible to study separate systems and discover that the mathematical structure needed to model them could be worked out explicitly, and that two of the axioms that are needed for the construction of standard quantum mechanics, called \emph{covering law} and \emph{weak modularity}, were violated by the situation of two separate systems. In other words, it became clear through these rigorous results that standard quantum mechanics is incapable of modeling two separate quantum entities due to a fundamental mathematical structural inability of Hilbert space to do so. 

Of course, there is an intimate relationship between this investigation on the axiomatics of separate quantum systems and the EPR paradox situation, and similar experimental situations that were explored by Einstein over the years in an attempt to show, precisely, that there is a problem in quantum mechanics when trying to describe separate systems. The problem cannot be solved by complementing the theory with additional variables for the quantum states, to allow for instance position and momentum to have simultaneous well-defined values and escape the limitation of Heisenberg's uncertainty relations. Anyway, this hidden-variables program proved in the end to be inconclusive, because of the obstructions of the so-called \emph{no-go theorems}, severely limiting the permissible hidden-variables theories \citep{neumann1932,bell1966,gleason1957,jauchpiron1963,kochenspecker1967,gudder1970}. In other words, the incompleteness evidenced by EPR-like reasonings was not an incompleteness in the sense of the hidden-variables to be associated with the quantum states, but an incompleteness in the sense of an incapacity of modeling separate quantum entities. 

Einstein's attempts to shed light on the problem of quantum non-separability were always expressed at a purely logical level, and in fact without any relation to subsequent experimental results, like those conducted in later years on entangled systems violating Bell's inequalities \citep{aspect1982,aspect1999,hensen2015}. For although thanks to these experimental results we know that it is possible to keep microscopic entities connected despite spatial separation, and this undoubtedly proves that Einstein's prejudice that spatial separation produces disconnection is unfounded, the experimental violation of Bell's inequalities never constituted an invalidation of EPR's \emph{ex absurdum} reasoning. This is because this reasoning, as Einstein has strongly tried to point out over the years (dissatisfied with how this aspect had remained hidden in the EPR article written by Podolsky), applies to a completely different experimental situation: that of two separate systems and not that of two systems in an entangled state. But two separate systems is exactly what quantum mechanics is structurally incapable of describing. This incapability is yes due to the existence of superposition states, as entangled states are, but as we have already pointed out it does not manifest itself in the quantum formalism at the level of states, but of the properties and measurements that are missing. And as we are going to explain, one of the two failing axioms in the study of separate quantum systems, the covering law, is in fact the equivalent of the \emph{superposition principle}. 

This is what we will try to illustrate in this article, presenting in a comprehensive but nonetheless non-technical way the results that led to the demonstration of the incompleteness of standard quantum mechanics due to its inability to describe separate systems. More precisely, after defining the notion of  \emph{separate systems} in Section~\ref{ss}, we introduce, in Section \ref{spf}, the key notions used in the Geneva-Brussels' formalism, and its first axiom. In Section~\ref{cdp}, we address the important question of the testability of conjunction and disjunction properties, and in Section~\ref{cqd} we analyze the difference between classical and quantum disjunctions. In Section \ref{ssr}, we briefly introduce the notion of \emph{superselection rule}, and in Sections~\ref{ortho} and \ref{atom}, we explain the second, third and fourth axioms, \emph{orthocomplementation}, \emph{state determination} and \emph{atomicity}. 
In Section~\ref{cwp}, we introduce the additional three more technical axioms,  \emph{covering law}, \emph{weak modularity} and \emph{plane transitivity}, needed to achieve a complete axiomatic representation of the standard Hilbertian formalism, that is, a lattice of properties that becomes a lattice of quantum properties, representable as the set of closed subspaces of a standard Hilbert space. We then conclude our axiomatic exposition by emphasizing, in Section~\ref{va}, that two of these seven axioms are violated by separate systems, proving the structural shortcoming of quantum mechanics that Einstein had foreshadowed without being able to clearly exhibit the reasons for such completeness, that is, the elements of reality that were missing in the theory. 

In Section~\ref{shl}, we use the powerful analogy between quantum behaviour and cognitive behaviour to shed further light on the issue of separability. The analogy was first advanced in the development of the domain called \emph{quantum cognition}, where a very effective modeling of the human cognitive domain was achieved through quantum mathematics \citep{aertsaerts1995,aertsbroekaertsmets1999,gaboraaerts2002,aertsczachor2004,khrennikov1999,altmanspacher2002}. But in a surprising reversal of perspective, it is also possible to observe that the behavior of microscopic systems becomes decidedly more intelligible when interpreted as the behavior of entities carrying meaning (obviously, not of the human kind). In other words, it is possible to speculate that the entities of the microscopic world possess a very similar nature to the entities of the human conceptual world, a hypothesis that has been explored in depth by our group in what has been named the \emph{conceptuality interpretation of quantum mechanics} \citep{aerts2009b,aertsetal2020}. The question we will ask is whether the inability of the quantum formalism to describe separate systems is really a limitation, or instead a reflection of the fact that separate systems would not really exist in reality. We will try to provide some insights into this issue, which we will show to be a difficult one, in this final section, by asking whether or not separate entities could also exist in the universe of human concepts. 

We conclude our article with Section~\ref{con}, in which we summarize what we have achieved.

\section{Separate systems\label{ss}}

We start by defining the notion of \emph{separate systems}. Note that when we use the term ``system’' (or ``entity'', we will use these two terms equivalently) we are referring to an idealization, namely, the possibility of describing and experiencing only a part of our total reality, without at the same time being forced to describe and experience phenomena external to that system. When we do so, we assign states and properties to that part and analyze how measurements on it behave. Of course, each system can have a corresponding external reality, and generally speaking, if a system $S_2$ is contained in this external reality of a system $S_1$, these two systems are separate. 

Consider the Earth and the Moon as a simple example drawn from our macroscopic world, as we conceive of them in the intuitive version of our reality. As they are modeled in classical mechanics, they are two separate systems. This does not mean, however, that they cannot dynamically influence each other through gravitation. When two systems are separate, it means that the idealization of being able to assign states to both and define measurements that act on these states, changing them into other states, is valid. Thus, measurements on one system should not need to take into account measurements possibly made on the other system to define their action on the states or, equivalently, should not be directly affected by the measurements on the other system. 

More precisely, if $S_1$ and $S_2$ are separate systems, then measurements performed on the former cannot depend, in an ontological sense, on measurements performed on the latter (simultaneously or in different moments), although, as we said, can nevertheless influence each other in a dynamical sense, for instance because both systems may interact by means of a given force field. In other words, quoting from \citet{aerts1984}: ``In general there is an interaction between separate systems and by means of this interaction the dynamical change of the state of one system is influenced by the dynamical change of the state of the other system. In classical mechanics for example almost all two body problems are problems of separate bodies (e.g., the Kepler problem). Two systems are non-separate if an experiment on one system changes the state of the other system. For two classical bodies this is for example the case when they are connected by a rigid rod.’’

In a general formalism that is operationally founded, like the one that was studied over the years by the Geneva-Brussels school, this condition of `separation of measurements', for measurements performed on two different systems, which in turn allow to define the notion of `separate systems', can easily be expressed. One can then construct the set of measurements associated with a composite system formed by two separate sub-systems, make its mathematical structure explicit and analyze it from an axiomatic viewpoint, that is, considering its compatibility with the axioms of the theory. The result of this analysis is that among the \emph{seven axioms} that reduce the `Geneva-Brussels operational formalism' to standard quantum mechanics in a generalized Hilbert space, two of them are never satisfied for a composite system formed by two separate quantum systems. More precisely, if a composite system satisfies the first four 
axioms plus the seventh one, and it also satisfies the fifth axiom, or the sixth axiom, then one of the two composing systems must be a classical system  \citep{aerts1981,aerts1982,aerts1983}. In other words, only the combination of a classical entity with a quantum entity, or the combination of two classical entities, can satisfy all the axioms, hence the combination of two quantum entities that are operationally defined to be separate cannot be described by quantum mechanics.\footnote{We should mention that the Geneva-Brussels operational formalism is general enough to describe both classical mechanics in a phase space and quantum mechanics in a Hilbert space, and likewise the combination of both possibilities in a structure technically called a `direct union over phase space of quantum components’, a state space structure that was used in Geneva to describe the so-called possibly continuous `superselection rules’ \citep{aertsdeses2002}; see also Section~\ref{ssr}.} 

This is the remarkable shortcoming of the Hilbert space formalism, to which we believe Einstein's intuition was pointing, but on which he could not have a rigorous understanding, since the mathematics that allows the separability problem to be brought into full view was not available in his time. To tell the truth, even today most physicists do not know  the mathematical formalism that allows the aforementioned result to be expressed and demonstrated, which explains why it is still so little known \citep{sassoli2019}, although these results are also extensively described in the The Stanford Encyclopedia of Philosophy \citep{Wilce2002}.

To fully understand this result it is certainly necessary to master the aforementioned axiomatic formalism developed by the Geneva-Brussels school. However, what we propose to do, in what follows, is to offer sufficient information on the content of the seven axioms that underlie the construction of the Hilbertian quantum structure, avoiding, on the one hand, to introduce the corresponding mathematical formalism (with a few exceptions, for the sake of clarity), since as we said it is not part of the standard baggage of a physicist, and, on the other hand, trying to remain sufficiently precise and comprehensible in our exposition. This attempt, to be comprehensive and at the same time avoid getting into the overly technical aspects, is undoubtedly one of the added values of the present article.

The first four of these seven axioms are very general and are connected with basic principles that one can expect to be satisfied for any type of system. On the other hand, the two ``failing axioms’’, those that are not satisfied by separate quantum systems are among the three axioms that had to be explicitly introduced to have the representation of the state space of quantum entities to become a standard Hilbert space. Different from the other four, they are  rather technical mathematical axioms with no straightforward connection to physical intuition or known physical principles. 
That the two failing axioms are of a technical mathematical nature while the non-failing axioms are very general, e.g., similar to the axioms for set theory, is an important aspect in relation to the following question: Is Hilbert space too specific a mathematical structure to describe all the reality associated with quantum systems, or, simply, are there no separate quantum entities? We will ponder on this question in the last part of this article.

\section{State, properties and the first axiom: completeness\label{spf}}

We start by introducing the key notions that are used in the Geneva-Brussels' formalism. Being an operational formalism, its building blocks are founded on what happens in the laboratory when measurements are executed on a given system. The approach is not only operational but also realist, in the sense that a physical system is considered to exist independent of whether it is measured upon or not, this being expressed by affirming that it `is’ always in one of its possible \emph{states}.

The archetypical operational situation consists of a system in a given state on which a measurement is performed, and a fundamental typology of measurements are those for which the corresponding experiments only have two outcomes, usually denoted `yes’ and `no’, as such yes-no measurements are those that can be used to define \emph{properties}. More precisely, whenever a system is in a state such that the `yes’ outcome of a yes-no measurement can be predicted with certainty, that is, with probability equal to $1$, one can say that the property operationally defined by such measurement is \emph{actual} for the system in that state. 

The set of properties defined in this way is equipped with a natural \emph{partial order relation}. More precisely, a property $a$ is said to be \emph{stronger than} 
a property $b$ if, whenever  the system is in a state such that property $a$ is actual, then also the less strong property $b$ is actual, for the system in question. This relationship, between $a$ and $b$, is denoted $a < b$.\footnote{In the literature on quantum axiomatics this is one of the most commonly used symbols to denote the `stronger than' relation. In articles by mathematicians it is usually denoted by the symbol $\le$, to make it clear that `being equal' is also a possibility. The choice of this symbol is related to the fact that in Hilbert space a property is usually represented by a closed subspace. If we consider two properties, with one stronger than the other, this corresponds in Hilbert space to the closed subspace representing the stronger property being contained in the closed subspace representing the weaker one, so the symbol $<$ acquires in this context the meaning of `being contained in' (in set theory one then uses the symbol $\subset$). To help physical intuition, particularly of readers unfamiliar with quantum axiomatics, we mention that the `partial order relation' can also be thought of as an an `implication relation'.} 

Let us give a simple example of the `stronger than’ relation. Take a piece of wood and assume it is in a state such that the property `being dry’ is actual. Then, it is easy to understand that the property `being burnable’ is also actual, and we can say that the dryness properties is stronger than the burnability property, for the system in that state. Note that properties are represented by \emph{orthogonal projection operators} in the quantum formalism. For two properties $a$ and $b$, associated with the projections $P_a$ and $P_b$, respectively, the relation $a < b$ is then equivalent to requiring $P_aP_b=P_bP_a=P_a$.

Let us also consider an example of a property that is \emph{not} stronger than another property. Always in relation to the piece of wood, which we will assume to be of a European type, let us suppose it is in a  `wet’ state. Then, consider the `floatability’ property, defined by the yes-no measurement consisting in putting the piece of wood on water to verify whether it floats (outcome `yes’) or does not float (outcome `no’). And consider also the `burnability’ property, defined by the yes-no measurement consisting of putting the piece of wood in contact with fire to verify whether it burns (outcome `yes’) or does not burn (outcome `no’). Clearly, when the wood-entity is in the `wet’ state, the `floatability’ property is actual but the `burnability’ property is not. Hence, floatability is not stronger than burnability, because the wet state exists. 

Having defined these basic notions, we can now explain the first axiom. It consists in requiring that the partial order relation we introduced, on the set of properties, defines a so-called \emph{complete lattice}, i.e., a \emph{complete partially ordered set}. This means that for each subset of properties there always exists an infimum and a supremum.

Let us start considering the notion of infimum. In the mathematical language where partial order structures are studied, a conjunction property,\footnote{In the technical literature, the conjunction properties are usually denoted \emph{meet properties}, see for instance \citet{aerts1999}.} usually noted 
$a\wedge b\wedge c\cdots$,  formed by the conjunction ($\wedge $) of a given subset of properties $\{a,b,c,\dots\}$, is precisely an infimum for such subset, and such a conjunction property exists for every subset of the set of properties of a system. To illustrate what this means, let us use the example of the piece of wood we have previously introduced. Suppose we consider the  `floatability' and `burnability' properties. Their conjunction corresponds then to the `floatability {\it and} burnability’ property. According to the meaning of the logical connective `and', this is a property which is clearly stronger than both properties which it is formed of, as a conjunction. 

In general, one can show that given a set of properties  $\{a_1,a_2,a_3,\dots\}$, the conjunction property $\wedge_i a_i$ is stronger than all of them, that is, $\wedge_i a_i < a_i$, for all $i$, and if there is a property $b$ that is stronger than all the properties in the set, that is, $b < a_i$, for all $i$, we also have $b <  \wedge_i a_i$. In other words, $\wedge_i a_i$ is the weakest property among the stronger ones, hence an infimum. This is a mathematical result, but still intuitive to understand. To return to our example, if a property is stronger than the two properties of `floatability’ and `burnability’, considered separately, say the property `being an European wood', it will also be stronger than the `floatability {\it and} burnability’ conjunction property. 

For the lattice to be complete, we need both the existence of an infimum and a supremum. As it is the case for the conjunction, the disjunction of a subset of properties can always be considered to be a property. In that respect, we can observe that a partially ordered set of properties associated to a given system always has a weakest property, which is the property `the system exists’. By definition, it is a property that is actual for any state of the system, as a consequence of the fundamental principle stating that if a system is in a given state, then it necessarily exists, and vice versa. This means that every property is stronger than this maximal property, and when a partially ordered set has a weakest property, and for all subsets an infimum exists, one can easily prove that for all subsets also a supremum exists, which can be expressed as an infimum, namely the infimum of all properties for which each property of the subset is stronger (the strongest property among the weakest). More precisely, given a set of properties $\{a_1,a_2,a_3,\dots\}$, we have seen that the associated infimum is $\wedge_i a_i$, and one can show that the associated supremum, expressed as an infimum, is given by the conjunction of all properties $b$ such that $a_i <  b$, for all $i$, i.e., in mathematical notation, by $\wedge_{a_j <  b} b$ \citep{aerts1982}. 

It is important to note that one cannot always find a yes-no measurement able to test the disjunction of a subset of properties, hence it will not always be an operationally defined property, whereas this is always the case for a conjunction property (see the next section). More precisely, when we only consider testable properties, the above special type of supremum is indeed equivalent to a logical disjunction property, but in general it will not be the case. 

Note also that, although we have described the completeness of the partially ordered set of properties, for the `stronger than’ partial order relation, as the first axiom, in fact, as we have seen, in the Geneva-Brussels formalism it is a theorem. The reason why it is generally mentioned as an axiom is because, in addition to the Geneva-Brussels formalism, these notions are also of interest to the broader field of \emph{quantum logic},\footnote{Quantum logic is a formal logical system that considers propositions as subspaces of a Hilbert space. This results in a specific calculus of propositions different from classical logic, in that some well-known principles, like the law of distributivity, no longer apply. Quantum logic can nevertheless be considered a genuine logic if we take logic as a tool for structuring reasoning within a specific domain. From this point of view, it provides a coherent system for reasoning about quantum entities, just as classical logic provides a coherent system for reasoning about macroscopic and deterministic entities.} for which, in the absence of an operational approach, the completeness of the lattice of properties is necessarily to be presented as an axiom.

\section{Are conjunction and disjunction properties testable?\label{cdp}}

Before continuing with our exposition of the subsequent axioms, it is important to ask whether conjunction and disjunction properties are operationally defined. In other words, can we identify yes-no measurements which can be used to test conjunction and disjunction properties? The answer is affirmative for the former and negative for the latter. 

Let us consider again the simple example we introduced, of the European piece of wood, and let us wonder how we would test the conjunction property of `floatability {\it and} burnability'. We know that both properties of `floatability' and `burnability' are actual if the piece of wood is in the `dry' state. But what makes us know this? Is there a measurement that can be used to test the `floatability {\it and} burnability' conjunction property, from which our knowledge would originate?

When a person ponders this question, s/he usually comes to the conclusion that such measurement would require to jointly execute both yes-no measurements, i.e., the one that tests the floatability and the one that tests the burnability. But, if that were true, we would be in trouble because, if we first test the burnability of the piece of wood, this will transform it into ashes, so it will not float any longer, hence the subsequent floating test will not give the outcome `yes’. If, on the other hand, we first check the floatability of the piece of wood, this will turn it into a wet piece of wood, hence it will no longer burn, and the burnability test will fail, i.e., it will give the outcome `no’. So, this cannot be the way to proceed to test the conjunction property of `floatability {\it and} burnability’, as we cannot simultaneously set fire to the piece of wood and immerse it in water, the two procedures being \emph{incompatible}. It is however not necessary to do so, as for testing a conjunction property we have to consider the opposite of a {\it conjunction experiment}, namely, a {\it disjunction experiment}.\footnote{In the dedicated literature, one does not speak of {\it disjunction experiment} but rather of \emph{product test}. We decided to use here  this terminology because it allows to explain more intuitively the content of these notions for the reader unfamiliar with the mathematics which they refer to. However, there would be also a more fundamental reason for adopting this terminology. Indeed, it is the disjunction that appears at the level of the structure of the formalism, rather than the somewhat more ad hoc definition of a product. For example, if one considers measurements with more than two outcomes, the disjunction measurement of two measurements will have, as outcome set, the set theoretic union, or join, of the set of outcomes of these two measurements.} Indeed, if we reflect carefully, we can observe that our knowledge that the considered conjunction of properties is actual for a dry piece of wood comes from the fact that we can freely and separately test  the `floatability' property or the `burnability' property, and whatever choice we make the `yes’ outcome will be obtained with certainty. 

Note that the disjunction of any subset of experiments always exists, as we just have to pick one and perform the corresponding test, hence a conjunction property expressing the conjunction of a given subset of properties also always exists and is operationally defined by means of a disjunction experiment. Of course, one needs for this that \emph{free choice} is part of our reality, that is, the possibility of genuinely selecting among different possibilities. In the example of the piece of wood, if we have the possibility of testing the `floatability' property or the `burnability' property, and the experimental protocol allows a free choice between these two possibilities, then what we are testing is truly the conjunction of the two properties, that is, the `floatability {\it and} burnability’ conjunction property \citep{aerts1994}.

Let us now consider the case of a logical disjunction property $a \vee b \vee c\cdots$, formed by the disjunction ($\vee$) of a given subset of properties $\{a,b,c,\dots\}$. If it also always existed and were operationally founded, it would clearly be the strongest among the weakest property, that is, a supremum for the considered subset. But, to test a disjunction property we would need a conjunction experiment, and we just observed in the piece of wood example that a conjunction experiment does not in general exist. This because experiments can change the state of the measured entity, so that there is no guarantee that two experiments can be jointly executed and, if executed sequentially, that the result would not depend on the chosen order. Or even worse, that after performing the first experiment the second would no longer be executable because the first has destroyed the measured entity. In other words, it is the nature of reality itself which prevents in general the existence of conjunction experiments, hence the testability of a disjunction of properties. 

So, going back to the notion of supremum, even though it can be operationally defined as a testable very large conjunction property (the stronger property among the weakest), it has a priori no relation at all with the logical disjunction property, the latter being in general not testable, hence not operationally definable.

\section{Classical and quantum disjunctions\label{cqd}}

The central topic of this paper being the non-separability of quantum systems, it is important to highlight a fundamental aspect of disjunction properties, when the corresponding conjunction measurements do not exist, hence the properties cannot be defined in operational terms. The obvious definition for a disjunction property, thinking about how we use the disjunction connective `or' in logic, would be the following: it is the property that is actual if and only if at least one of the properties in the disjunction is actual. But is that always true?

To answer this question, suppose the disjunction in question is only about two properties,  $a \vee b$, operationally defined by two yes-no measurements. Let us further assume that correlations occur when jointly executing these tests, like those in EPR-paradox situations, when Bell’s inequalities are violated. In other words, the two properties we are considering, in the disjunction $a \vee b$, are those typically tested by Alice and Bob in this kind of setting. For example, the property of being up along a given spatial direction for spin-$1/2$ entities, when the tests are conducted with Stern-Gerlach apparatuses, when Alice and Bob are jointly experimenting with a composite system formed by two spin-$1/2$ entities in an entangled \emph{singlet state}. 

Due to the presence of, \emph{entanglement},\footnote{In quantum mechanics, entanglement is a situation in which a composite system is in a state that contains no information about the (pure) states of its subsystems. This is mathematically expressed as a non-factorizability of the state of the composite system, being a superposition involving both subsystems, at the origin of quantum correlations.} the conjunction measurement consisting of two yes-no measurements, $a$ and $b$, jointly performed by Alice and Bob, will produce anti-correlations, if both chose the same spatial direction, i.e., the experiment will give with certainty one of following two outcomes: (yes, no) or (no, yes), and both are of course equally possible. Indeed, if Alice’s spin is observed to be `up', along a given direction, Bob’s spin is observed to be `down', along that same direction, and vice versa, with the `up' and `down' outcomes playing here the role of the `yes' and `no' answers. Now, the logical way to define a disjunction property using a conjunction measurement is to consider that the property is \emph{actual} (answer `yes') if Alice's test, or Bob's test, or both, give the answer `yes', and \emph{potential} otherwise. In other words, the three outcomes (yes, yes), (yes, no) and (no, yes) are associated to the answer `yes', for the yes-no measurement testing the disjunction property $a \vee b$, and the  outcome (no, no) is associated with the answer `no'. 

However, in a situation of entanglement, since perfect anti-correlations are observed when Alice and Bob orient their instruments along the same direction, the (yes, yes) and (no, no) outcomes are excluded. This means that $a \vee b$ is always actual, without the need for one of its component properties, $a$ and $b$,  to also be actual. Indeed, we can predict with certainty the `yes' outcome of the test for the disjunction property, but we cannot predict with certainty the `yes' outcome of Alice’s or Bob’s individual tests, the composite system being in an \emph{entangled state} and not in a so-called \emph{product state}. 

So, disjunction properties behave in this unexpected way in quantum mechanics not because a conjunction measurement would not be available, being clear that in Bell-test experimental situations Alice and Bob can jointly perform their tests, as the associated observables commute, hence are assumed to be compatible.\footnote{Note however that this assumption may not always be true, as evidenced by the fact that in many experimental situations the so-called `no-signaling conditions' are violated \citep{aabgssv2019}.} The fact that in quantum logic the disjunction ($\vee$) cannot be interpreted as in classical logic is just because there are states in the formalism that give rise to Bell-type anti-correlations \citep{aertshndtgabora2000}.

From the perspective of Hilbertian formalism, this difference between a quantum disjunction property and a classical disjunction property can easily be observed in the fact that  yes-no measurements are represented by \emph{orthogonal projection operators}, and measurements with more than two outcomes are more generally represented by \emph{self-adjoint operators}. What we explained then translates in the standard quantum formalism in the fact that if one takes a state-vector in the closed subspace corresponding to the orthogonal projection operator representing one of two yes-no measurements, and takes another state-vector in the closed subspace corresponding to the other projection operator, i.e., to the other yes-no measurement, one can always consider a linear combination of these two vectors, representing a so-called \emph{superposition state}, such that only the perfectly anti-correlated (yes, no) or (no, yes) outcomes will be possible for the self-adjoint operator representing the conjunction of these two yes-no measurements, typically described by the tensor product of their observables. On the other hand, if the composite system in question is in a state such that entanglement correlations are absent, that is, it is in a \emph{product state}, we are in a situation where the disjunction property behaves as classically expected, since it can then be proven that the disjunction property is actual if and only if one of the constituting properties is actual \citep{aerts1981,aerts1981b}.

\section{Superselection rules\label{ssr}}

Before proceeding with the explanation of the second axiom, it is a good time to also introduce the notion of \emph{superselection rule} (ssr), which we will need later. Superselection rules were historically introduced in quantum mechanics as a way of generally expressing restrictions on the nature and scope of possible measurements \citep{Wick1952}. To understand what a superselection rule implies for properties, let us observe again that the conjunction property $a \wedge b$ is actual if and only if $a$ is actual and $b$ is actual, hence  $a \wedge b$ is the property `$a$ \emph{and} $b$'. We have seen that the disjunction property  $a \vee b$ is defined in terms of a large conjunction of properties:  $a \vee b = \wedge_{a,b <  c} c$. Hence, a fundamental difference between $a \wedge b$ and $a \vee b$ is that the former is defined only in terms of $a$ and $b$, whereas the latter depends on all the other properties of the system in question, which means that if we add more properties to the system, this will affect $a \vee b$. We have also seen that even when $a$ and $b$ can be jointly tested, it is not necessarily the case that the actuality of $a \vee b$ implies the actuality of $a$ or the actuality of $b$. As we have explained, the reason is the existence of superposition states that produce correlations. 

The existence of a superselection rule is precisely a situation where such a situation does not arise, hence a superselection rule constitutes an inhibition of the superposition principle. More precisely, one says that $a$ and $b$ are separate by a superselection rule ($a$ ssr $b$) if for every state  such that $a \vee b$ is actual in that state, then either $a$ is actual or $b$ is actual. When $a$ ssr $b$, it can be easily proven that this is also the situation where $a \vee b$ does not depend anymore on all the other properties of the system, but just on $a$ and $b$, as it is the case in classical mechanics.  

The notion of superselection rule is important for our discussion because if $a$ and $b$ are two \emph{separate properties} of a system, then $a$ and $b$ are separated by a superselection rule. By separate properties we mean properties that can be tested by  \emph{separate yes-no measurements}, that is, by measurements  that can be performed together without influencing each others.

Note also that the existence of a superselection rule between $a$ and $b$ is due to the existence of a yes-no measurement that can test whether $a$ is actual or $b$ is actual. This means that in quantum mechanics the \emph{superposition principle} emerges from a scarcity of measurements, i.e., from the impossibility, for two properties $a$ and $b$, to construct a measuring apparatus capable of testing whether $a$
is actual or $b$ is actual. Since for macroscopic systems we are in principle always able to construct such an apparatus, this explains why the superposition of their states is not observed.

\section{The second axiom: orthocomplementation\label{ortho}}

Let us now come to the description of the second axiom, which asks for the lattice of properties the existence of an \emph{orthocomplementation}, which is a quantum-like generalization of what a \emph{complement} represents in classical physics in phase space.\footnote{To aid the intuition of readers unfamiliar with quantum axiomatics, we mention that an orthocomplementation  can also be understood as a quantum-like generalization of what a \emph{negation} is in classical logic.} More precisely, the orthocomplement of a property $a$, usually denoted $a'$, is also a property and, similarly to the \emph{complement} in set theory must obey the rule:\footnote{In classical logic, this is the equivalent of the so-called \emph{modus tollens}.}  $a <  b$ ($a$ is stronger than $b$) implies that $b'<  a'$ ($b'$, the orthocomplement of $b$, is stronger than $a'$, the orthocomplement of $a$). In addition to that, we also need to have the other rules of set theory that the complement obeys, i.e., $(a')'=a$ and $a\wedge a'=0$, where $0$ denotes the equivalence class of all yes-no measurements for which the outcome is never with certainty `yes’, called the \emph{minimal property}. 

In the Hilbertian formalism, properties $a$ and $b$ are associated with orthogonal projection operators $P_a$ and $P_b$, respectively, and the corresponding orthocomplements $a'$ and $b'$ can be defined by considering the orthogonal projection operators $P_{a'}=I-P_a$ and $P_{b'}=I-P_b$, respectively, where $I$ is the unit operator. To see this, we recall that the relation $a <  b$ is the equivalent of requiring $P_aP_b=P_bP_a=P_a$. So, the relation $b'<  a'$ is the equivalent of requiring $P_{b'}P_{a'}=P_{a'}P_{b'}=P_{b'}$. This is clearly satisfied since 
\begin{equation}
P_{b'}P_{a'}=(I-P_b)(I-P_a)=I-P_a-P_b+P_bP_a= I-P_a-P_b+P_a=I-P_b=P_{b'},
\end{equation}
and similarly for the relation $P_{a'}P_{b'}=P_{b'}$. Also, $(a')'=a$ corresponds to 
\begin{equation}
I-P_{a'}=I-(I-P_a)=P_a.
\end{equation}
Hence the equality is obeyed, and finally  $a\wedge a'=0$ corresponds to 
\begin{equation}
P_aP_{a'}=P_a(I-P_a)=P_a-P_a^2=P_a-P_a=0,
\end{equation}
where for the last equality we have used the idempotency property of orthogonal projections, and $0$ is here the zero projection, describing the minimal property of the system. So, the structure of a Hilbert space naturally  induces an orthocomplementation.

To explain the content of this second axiom, and to what extent it can be operationally interpreted, let us first consider the notion of \emph{orthogonality}. If one considers an arbitrary yes-no measurement, the default situation is the following. There are states of the system for which the measurement gives with certainty the `yes' outcome, there are states of the system for which the measurement gives with certainty the `no' outcome, and there are states of the system for which the outcome is indeterminate, in the sense that it can give either the `yes' outcome or the `no' outcome. 

Based on this observation, one can introduce a natural notion of orthogonality, in relation to the states of a system, in the following way: two states are orthogonal if and only if a yes-no measurement exists such that, when executed on a system in one of the two states, it would give with certainty the `yes' outcome, whereas it would give with certainty the `no’ outcome if that same system is in the other state. This orthogonality notion for the states can be applied also to properties: two properties are orthogonal if all pairs of states that make these properties actual are orthogonal states. And we can also define an orthogonality relation between properties and states: a state is orthogonal to a property if it is orthogonal to all states that make this property actual.  

We observed that the first axiom is a theorem when one adopts an operational approach. The second axiom is however different, since it is not possible to define operationally, in a general way, the orthocomplement of a property, as we can do instead for the orthogonal of a property. The reason is the very existence of the disjunction experiments used to operationally prove axiom 1. More precisely, consider an arbitrary property $a$. Its orthocomplement $a'$ is another property which is necessarily orthogonal ($\perp$) to the original property ($a'\perp a$). However, more is required for an orthogonal property to be also an orthocomplement property. In other words, orthocomplementation is more than just asking for orthogonality. 

As we can see using the specific Hilbertian formalism, orthocomplementation obeys the rule that the orthocomplement of a property stronger than another property is less strong than the orthocomplement of the latter, but this is not anymore the case if the properties are conjunction properties, that is, properties tested by yes-no disjunction measurements. To see why it is so, let us use again the example of the piece of wood. Earlier, we considered that it was a European wood, but now let us consider that the type of wood can be either European or Ebony (Ebony is a very dense non-European wood that does not float), depending on the state of the wood entity.  So, when it is in the state  `dry European wood', we know that the two yes-no measurements testing the `floatability' and  `burnability' properties, respectively, give both the answer `yes’ with certainty, which means that the `yes' answer is certain also for the yes-no disjunction measurement testing the `floatability \emph{and} burnability' conjunction property. On the other hand, when in the state `wet piece of Ebony wood’, these same yes-no measurements will give both the answer `no’. But if we reverse in these measurements the role of the `yes' and `no' outcomes (defining so-called \emph{inverse} yes-no measurements), then the `yes' answer will be certain for the yes-no disjunction measurement formed by these inverse measurements, now testing the `sinkability  \emph{and} fireproof' conjunction property. 

However, there are states of the piece of wood where the outcome of the disjunction test is fundamentally indeterminate, and will either give the answer `yes' or the answer `no' depending on which test is being executed. Indeed, suppose the piece of wood is in the state `dry piece of Ebony wood’. Then, the `yes' and `no' outcomes are both possible for the test of the `floatability \emph{and} burnability' conjunction property, and similarly for the piece of wood being in the state `wet piece of European wood’. But this is a problem for the orthocomplementation rule we mentioned above. Indeed, the `floatability \emph{and} burnability' conjunction property is `stronger than' the floatability property, hence, by changing the roles of the `yes’ and `no’ outcomes, we should have that the sinkability (non-floatability) property should be stronger than the  `sinkability \emph{and} fireproof' conjunction property. But this cannot be the case, since in the `dry piece of Ebony wood’ state the sinkability property is actual, whereas the   `sinkability \emph{and} fireproof' conjunction property is indeterminate, hence not actual. In other words, the above mentioned orthocomplementation rule fails. 

It is the presence of the disjunction measurements that does not allow to obtain an operational foundation for an orthocomplement property. What one needs to do is to introduce the notion of \emph{primitive} yes-no measurement (a type of test called \emph{ortho test} in \citet{aerts2009} is a good candidate for a primitive test), which is a measurement that is not of a disjunction kind. So, for the  lattice of properties to have an orthocomplementation, inherited by the orthogonality relation, we need to ask more: for each property there has to be a primitive yes-no measurement testing it, and this is the non-operational part of axiom 2. 

We can understand a primitive measurement as a measurement that eliminates as much as possible all non-equivalent components in the question it asks, so that, if initially viewed as a conjunction of measurements, all these components become equivalent as regards the question they ask. In a sense, an experimenter will always try, when conceiving her/his measurement processes, to ideally make them primitive interrogative processes, where a single non-decomposable question is posed to the measured system. 

To conclude this section, it is worth also mentioning the operational definition of a \emph{classical property}. A property is said to be classical if the yes-no measurement testing it gives either `yes' with certainty or `no' with certainty, for all possible states, that is, the indeterminate case does not occur. A disjunction measurement is clearly in general not of this kind, and in fact it can be proven that if a disjunction yes-no measurement is classical then all the yes-no measurements composing it must be also classical and equivalent \citep{aerts1983,aerts2009}, that is, they all test the same property. And this means that the notion of disjunction measurement, needed to operationally define conjunction properties, already introduce a non-classical (quantum-like) aspect into the formalism, in a very intrinsic way.

\section{The third and fourth axioms: state determination and atomicity\label{atom}}

The third axiom in the quantum logic axiomatic approach is called \emph{state determination}. It is the assumption that the state of a system, in a given moment, is always fully determined by the properties that are actual in that moment, and more specifically by their conjunction. In the early years of axiomatic development \citep{piron1964,aerts1981,aerts1982}, the duality between states and properties, which has only been considered in more recent years \citep{aerts2009}, was not yet explored and states were defined as the sets of all actual properties, usually described as their infimum, if the axiom of completeness was satisfied. This means that axiom 3 was a priori satisfied as a consequence of how states were defined. Only later was it realized that it required an axiom of its own and could not be derived from the operational content of the theory \citep{aertsValckenborgh2004}.

So, according to axiom 3, a state depends completely on the notion of property, and more precisely on those properties that are actual for the state in question. Undoubtedly, this introduces a constraint in a theory that obeys it. Indeed, as we said, in general terms one could ask the notions of state and property to be independent,  in the sense that two states can be thought of as different while actualizing exactly the same properties. This means that states could, in general, differ in their behavior in relation to \emph{potential} properties while defining the same set of actual properties. 

Instead, with axiom 3, the state space of the system is required to be such that it does not possess any freedom to vary a state without also varying the actual properties characterizing it. Awareness of the excessive limitation that this axiom imposes arose in the 1990s, when, on the one hand, the Brussels group investigated the categorical structure of the Geneva-Brussels approach \citep{aerts1999a,aertscolebunders1999,coeckestubbe1999,aertscolebunders2002} and, on the other hand, began the research in quantum cognition \citep{aertsaerts1995,aertsgabora2005}. Categorical structure research led to a version of the Geneva-Brussels formalism that very elegantly expresses a fundamental duality between states and properties, while optimally distinguishing between operationally provable structures and structures requiring axioms \citep{aerts2009}. Quantum cognition research made it clear that measurements are not only associated with properties, but are also always carried by \emph{contexts}, able to change states even deterministically in some situations.  This insight led us to add a fundamental element to axiomatics by introducing, in addition to the state and properties of an entity, the contexts of an entity, described in a way that is perfectly independent of states and properties. We called this context-enhanced axiomatic structure a \emph{state context property system}, or SCoP \citep{aertsgabora2005b}. 

The fourth axiom is the request of \emph{atomicity} of the lattice of properties.\footnote{For completeness, we point out that it can be proven that, the axiom of atomicity being satisfied, the lattice of properties satisfies a much stronger property, namely the one of \emph{atomisticity} \citep{aerts2009}.} It requires that the states of a system be the \emph{atoms} of the property lattice. This notion of atoms has of course nothing to do with those of physics. It is just a mathematical analogy characterizing the smallest element in a collection relative to the partial order relation considered, which in our case means the strongest element in a collection of properties. More precisely, an atom $a$ is an element of the lattice of properties which is different from $0$ and such that if there is a property $b$ for which $b <  a$ ($b$ is stronger than $a$), then either $b=0$ or $b=a$. In other words, atoms are the smallest elements of a lattice different from the minimal property  $0$.  

Once again, this notion of atomicity is quite natural if one considers that the state of a system  in a given moment is characterized, as per the axiom 3, by the properties that are actual in that moment, and that the conjunction of these properties is necessarily also actual and is the infimum of all the actual properties defining the state in question. Now, what axiom 4 requires is that such infimum, usually called \emph{property state}, is an atom of the system's property lattice. Note that an atom is necessarily a property state. To show this, suppose that $a$ is an atom. Let  $p$ be a state making $a$ actual. If we denote $s(p)$ the property state associated with $p$, we have $s(p) <  a$. But since $a$ is an atom, we necessarily also have $s(p) = a$. Hence, $a$ is a property state. But the reverse is generally not true: not all property states are necessarily also atoms, and it is the role of axiom 4 to insure that this is always the case, i.e., that there is full correspondence between the property states and the atoms of a system's property lattice. At the level of the complex Hilbert space of quantum mechanics, structures are studied that do not satisfy the axiom of atomicity, for example the Von Neumann algebras associated with the Hilbert space, where states are not necessarilyp smallest elements of the lattice of properties \citep{neumann1930}

Summarizing what we have explored so far, the first four axioms that are necessary to arrive at an Hilbertian representation are partly proved by adopting an operational approach, and those aspects that cannot receive an operational foundation nevertheless remain close to our intuition of what conditions a physical system should satisfy.

\section{Covering law, weak modularity and plane transitivity\label{cwp}}

Let us now consider the three axioms called \emph{covering law}, \emph{weak modularity} and \emph{plane transitivity}. They are what is needed for the lattice property structure to become representable into a \emph{Hilbert space}, and they were introduced as technical axioms, without being supported by any specific physical intuition. 

The fifth axiom, the covering law, demands that the supremum $a \vee b$ of an atom $a$ and a property $b$ \emph{covers} this property $b$, in the sense that there does not exist a property $c$, different from $b$ and different from $a \vee b$, in between $a$ and $b$. Or, formally, if $a$ is an atom and $b$ a property, such that $a\wedge b=0$ and $b <  c <   a \vee b$, we have either $b=c$ or $a \vee b=c$. Without going into the technical aspects, it can be shown that a complete atomic lattice satisfying the covering law will be isomorphic to a \emph{projective geometry}. Using the fundamental theorem of projective geometry, it is then possible to construct a \emph{vector space} coordinating this geometry and representing the lattice of properties\citep{piron1964}.\footnote{To obtain a vector space for the set of states, a collection of deep theorems of projective geometry needs to be used, rooted in contributions from Pappus of Alexandria, Johannes Kepler, Blaise Pascal, August Ferdinand M\"{o}bius, Henri Poincar\'e, and many other less famous mathematicians.} This means that the covering law axiom introduces the linearity of the set of states, hence, it is at the root of the linear structure of quantum mechanics or, to put it differently, it is about the \emph{superposition principle}.

Note that although the notion of superposition of states is intrinsically linked to the linearity of the Hilbert space, it is possible to introduce it on a more fundamental level in the Geneva-Brussels formalism. More precisely, considering two states $p$ and $q$, one says that a third state $r$ is a superposition of $p$ and $q$ if the intersection between the set of properties that are actual in $p$ and the set of properties that are actual in $q$ is contained in the set of properties that are actual in $r$. This intersection can of course be empty, which means that the superposition principle is connected with the existence of states that make a disjunction of properties $a\vee b$ actual, but not the component properties $a$ and $b$. 

Regarding the sixth axiom, weak modularity, what we can say is that, on the level of the Hilbert space, it is the equivalent of the \emph{Cauchy completeness} of that Hilbert space, if the formalism is represented within a real, complex or quaternionic vector space structure. In other words, it is the requirement that all converging sequences, in the Cauchy sense, converge within the state space and that there are no missing elements.\footnote{More precisely, Cauchy completeness ensures that ``gaps'' do not exist in the space in question, in the sense that Cauchy sequences always converge to elements in it. For example, the set of real numbers with the usual distance metric is Cauchy complete, but the set of rational numbers is not, as there are Cauchy sequences of rational numbers that converge to irrational numbers.} In technical terms, the axiom requires that if $a$ and $b$ are two properties such that $a <  b$, then $(a\vee b')\wedge b=a$. 

Another equivalent definition of weak modularity is to require that whenever for two properties, $a$ and $b$, we have $a < b$, then the sublattice generated by the two properties and their orthocomplements is a \emph{Boolean lattice}. In other words, whenever two properties are connected by the partial order relation, they generate a classical sublattice. Thus, weak modularity introduces some additional classicality into the lattice of properties. 

All these six axioms are satisfied by quantum mechanics and also by classical mechanics, and in a sense also the opposite is true: the six axioms force the `state property space' formalism to become equivalent to quantum mechanics with superselection rules. And if there are no superselection rules, we get ordinary quantum mechanics in a single Hilbert space. And if all states are separate by a superselection rule, we get classical mechanics. So, the formalism emerging from these six axioms is more general than quantum mechanics, in the sense that it introduces in a natural way the possibility of superselection rules. 

More precisely, limiting here our discussion to the situation without superselection rules, the six axioms we have described allow one to prove a representation theorem stating that the lattice of properties can be represented as the set of \emph{closed subspaces} of a \emph{generalized Hilbert space} \citep{piron1964,aerts2009}. Note, however, that such a generalized Hilbert space is not necessarily a vector space over the complex numbers, which is the Hilbert space of standard quantum mechanics. Other number fields are also possible for a generalized Hilbert space, for example the quaternions, and more exotic number fields are not excluded. So, we could say, standard quantum mechanics is not fully reached with only these six axioms.

The mathematical result that made it possible to arrive at a complete axiomatic representation of the standard formalism of quantum mechanics came only in 1995, with a theorem of Maria Pia Sol\`er \citep{soler1995}. Working on generalized Hilbert spaces (today better known as \emph{orthomodular spaces}) within the research field of \emph{abstract algebras}, she was able to show that if a generalized Hilbert space contains an infinite orthonormal basis, then it has to be isomorphic to one of the three standard Hilbert spaces: the real, the complex or the quaternionic.

To make contact with Sol\`er's result, and complete the axiomatics of standard quantum mechanics, a seventh axiom was therefore needed, to ensure that the lattice of properties would satisfy Sol\`er's condition. This is the plane transitivity axiom that was identified in 2000 \citep{aertssteirteghem2000}.\footnote{For finite-dimensional generalized Hilbert spaces the seven axioms are however still compatible with vector spaces over number fields that can be different from  the real, complex or quaternionic numbers.} The formulation of this seventh  axiom is very technical (and not particularly elegant) and cannot be easily formulated without the introduction of a number of sophisticated mathematical notions, hence, differently from the previous five, it will not be given here. The interested reader may refer to  \citet{aertssteirteghem2000,aerts2009}.

\section{The violated axioms\label{va}}

We have seen in the previous sections that both classical systems (a system with only classical properties)  and quantum systems can be described by a `state property space' formalism where the set of properties is a complete, state determined, atomic and orthocomplemented lattice that satisfies the covering law, is weakly modular and plane transitive. Axioms 1, 2, 3, 4 and 7 are fully compatible with the description of the most simple of all compositions of two systems, that of two separate systems, whereas axioms 5 and 6 are not. As we mentioned already, axiom 5 (covering law) is equivalent to the superposition principle, which means that separate systems are in direct conflict with it, in the sense that they call its general applicability into question. In what follows, without entering in the mathematical details, we briefly provide the conceptual steps needed to understand the reason of the violation of the fifth and sixth axioms. 

First, one observes that a classical system is such that, if two of its states are different, then they are also orthogonal. Hence, for a classical system the orthogonal relation becomes trivial. Next, one observes that if axioms 1, 2, 3 and 4 are satisfied, plus axiom 5 (or axiom 6) is satisfied, then if two states are separate by a superselection rule, they are necessarily orthogonal states. In other words, axiom 5 (or axiom 6) makes it impossible to have non-orthogonal states that are separate by a superselection rule. But these are precisely the states that exist for a composite system formed by separate sub-systems. Hence, if axiom 5 (or axiom 6) is satisfied, one cannot describe two separate quantum systems. What remains possible, however, is that one of the sub-systems, or both, are classical systems \citep{aerts1982}. The consequence of the above is that if one uses quantum mechanics to describe a composite entity formed by two quantum sub-entities, paradoxical situations will necessarily be encountered, as the formalism is structurally incapable to describe such a situation. More precisely, it cannot describe nonorthogonal states that are separate by a superselection rule. 

We can therefore assert that if quantum mechanics is incomplete it is not because it does not describe hidden variables pertaining to a finer description of the states, but because it is unable to describe separate physical systems. Of course, incompleteness is relevant only if separate systems do exist in nature, if they don't, at a fundamental level, then this structural limitation has no practical consequences. Before analyzing this possibility in the next section, we would  like to also illustrate to the reader a direct and constructive demonstration of the impossibility of the quantum formalism to describe a joint measurement formed by two separate measurements, because of the superposition principle. 

Suppose that $S$ is a system formed by two sub-systems $S_1$ and $S_2$, with Hilbert spaces ${\cal H}$, ${\cal H}_1$ and ${\cal H}_2$, respectively. If $S_1$ and $S_2$ are separate sub-systems, their measurements are separate measurements that do not influence each other. If ${\cal M}_1$ is a measurement on $S_1$ and ${\cal M}_2$ is a measurement on $S_2$, one can then define a combined measurement ${\cal M}$ such that: (1) the execution of ${\cal M}$ on $S$ corresponds to the execution of ${\cal M}_1$ on $S_1$ and of ${\cal M}_2$ on $S_2$; (2) the outcomes of ${\cal M}$ are given by all possible couples of outcomes obtained from ${\cal M}_1$ and ${\cal M}_2$. What we need then to do is to show that one cannot exhibit a self-adjoint operator $O$ able to represent ${\cal M}$. 

For this, consider two arbitrary projections $P_1^{I_1}$ and $P_2^{I_2}$, in the spectral decomposition of the self-adjoint operators $O_1$ and $O_2$ associated with measurements ${\cal M}_1$ and ${\cal M}_2$, respectively. Here $I_1$ and $I_2$ are subsets of the outcome sets $E_1$ and $E_2$ of the two measurements, respectively. Consider also the spectral projection $P_{AB}^{I_1\times I_2}$ of $O$, where $I_1\times I_2$ is the subset of the outcome sets of ${\cal M}$, formed by all couples $(x,y)$ of elements $x\in I_1$ and $y\in I_2$. Without going into the details, one can then show, as one would expect, that $[P_1^{I_1}, P_2^{I_2}]=0$, so that also $[O_1, O_2]=0$, and $P_{AB}^{I_1\times I_2}=P_1^{I_1}P_2^{I_2}$. The next step is to consider a state $\psi\in {\cal H}$ which is the superposition $\psi = {1\over\sqrt{2}}(\phi + \chi)$ of a state $\phi$, belonging to the subspace $P_1^{I_1}(\mathbb{I}-P_2^{I_2}){\cal H}$, and a state $\chi$, belonging to the subspace $(\mathbb{I}-P_1^{I_1})P_2^{I_2}{\cal H}$, orthogonal to the latter. It follows that: 
\begin{equation}
P_1^{I_1}\psi={1\over\sqrt{2}}\phi, \quad 
(\mathbb{I}-P_1^{I_1})\psi=
{1\over\sqrt{2}}\chi, \quad P_2^{I_2}\psi=
{1\over\sqrt{2}}\chi,\quad (\mathbb{I}-P_2^{I_2})\psi={1\over\sqrt{2}}\phi.
\end{equation} 
This means that when the bipartite system is in state $\psi$, there is at least two possible outcomes $x_1\in I_1$ and $x_2\in E_1- I_1$, for measurement ${\cal M}_1$, and at least two possible outcomes $y_1\in I_2$ and $y_2\in E_2- I_2$, for measurement ${\cal M}_2$. Therefore, the four outcomes $(x_1,y_1)\in I_1\times I_2$, $(x_1,y_2)\in I_1\times (E_2- I_2)$, $(x_2,y_1)\in (E_1- I_1)\times I_2$ and $(x_2,y_2)\in (E_1- I_1)\times(E_2- I_2)$ should be all possible outcomes of measurement ${\cal M}$, considered that we have assumed  ${\cal M}_1$ and ${\cal M}_2$ to be separate measurements. We have: 
\begin{equation}
P_{AB}^{I_1\times (E_2-I_2)}\psi=P_1^{I_1}(\mathbb{I}-P_2^{I_2})\psi=\phi, \quad P_{AB}^{(E_1-I_1)\times I_2}\psi=(\mathbb{I}-P_1^{I_1})P_2^{I_2}\psi=\chi,
\end{equation}
hence $(x_1,y_2)$ and $(x_2,y_1)$ are possible outcomes of ${\cal M}$. But we also have: 
\begin{equation}
P_{AB}^{I_1\times I_2}\psi=P_1^{I_1}P_2^{I_2}\psi=0, \quad P_{AB}^{(E_1-I_1)\times (E_2-I_2)}\psi=(\mathbb{I}-P_1^{I_1})(\mathbb{I}-P_2^{I_2})\psi=0,
\end{equation}
 hence $(x_1,y_1)$ and $(x_2,y_2)$ are not possible outcomes of ${\cal M}$, which means that ${\cal M}_1$ and ${\cal M}_2$ are not separate measurements. In other words, because of the superposition principle, a joint measurement ${\cal M}$, formed by two separate measurements ${\cal M}_1$ and ${\cal M}_2$, cannot be consistently described in standard quantum mechanics \citep{aerts1982,aerts1984b}. 

Note the importance of working out the above proof without introducing a specific tensorial representation ${\cal H}={\cal H}_1\otimes {\cal H}_2$, so that one can be certain that its conclusion is inescapable. Again, we see that the shortcoming of quantum mechanics in describing separate systems cannot be identified at the level of the states, as in a sense there is an overabundance of them, but at the level of the properties, described by orthogonal projection operators. This overabundance of states is of course connected with the deficiency of properties and measurements, in the sense that certain properties of a bipartite system formed by separate components cannot be represented by orthogonal projection operators. Therefore, neither can the experiments capable of testing them be defined. 

We also want to mention how the impossibility of describing separate systems manifests itself if we keep in mind how the fundamental theorems of projective geometry plays a key role in Piron's representation theorem \citep{piron1964}. As we already noted, it is the covering law that makes the lattice of properties conceivable as a projective geometry, with the points corresponding to the atoms (state properties) of the lattice of properties and the lines of the projective geometry corresponding to the planes, consisting of the joins of pairs of atoms. But one of the axioms of projective geometry requires each line to contain at least three points \citep{coxeter1994}, and that is precisely the axiom that goes wrong for separated systems. 

Indeed, for separated systems there exist planes consisting of only two points. More precisely, the planes consisting of the joins of atoms corresponding to products of state properties are such planes, containing only two points, i.e., the two atoms that are products of state properties themselves, as a consequence of the existence of superselection rules between them. Since these products of state properties are not necessarily orthogonal (at least one of the elements in a product needs to contain orthogonal states for this to be the case) such superselection rules cannot be described in a formalism where the covering law is valid. Within the corresponding projective geometry this is the equivalent of a plane needing to contain at least three points. So, no projective geometry, as an intermediate step towards the construction of a vector space, is possible for the situation of separate systems. 

One may wonder why, generalizing the notion of projective geometry, the requirement that a line contains at least three points has been made part of the definition and sets of two points were not considered lines. It would be interesting to know a general answer to this question starting from the fundamental notion of \emph{geometry}. A related question one may ask is whether another geometric or algebraic structure, that spontaneously contains the situation of separate systems, has been studied by mathematicians. In partial answer to this question, we can say that the structure of a \emph{complete orthocomplemented atomic lattice}, which is also the structure of the property lattice of a physical entity for which the first four axioms are valid, is probably the appropriate one to investigate, and there is some work on this, although not much; see \citet{maclaren1964,parrocchia2023} and references therein. The categorical connections we discovered with closure spaces is also important in this regard, being that there is more abundance of mathematical results for these structures
\citep{faure1994,aertsdhooghesioen2011}. 

Purely operational axiomatic work
probably lies ahead as well. Which structure can be identified as operationally inferable and therefore requires less structure to be added from an axiomatic standpoint is not a fully resolved question at present, as it is the case with which classical structures can be identified as substructures, operationally or axiomatically. An example is how we managed to operationally
introduce the notion of `topological' (considering
topological as the classical for closure) as an alternative to `classical' (in the sense of deterministic) and characterize the corresponding substructures in citet{aertsdhooghesioen2011}, with the intention of making a pass towards the generally continuous
(hence topological) formalisms of classical mechanics.

We therefore invite the mathematical community to renew the study of complete orthocomplemented atomic lattices and their corresponding closure spaces, to understand whether within them lies the possibility of a coordination admitting examples having an computational operability and specificity similar to that of a complex Hilbert space.

\section{Separability in human language \label{shl}}

We conclude our analysis by addressing the following question.

\emph{Should we consider Hilbert space as too special a space to describe physical systems in all generality?} 

Considering its structural inability to describe separate systems, the answer would seem to be yes. However, it is also possible that what seems to be a limitation of the formalism is in fact not such, because perhaps in physical reality there are no genuine separate systems and everything is inexorably connected to everything else, in a perfectly quantum coherent way. 

One way to explore this question is to exploit the powerful analogy that exists between, on the one hand, microscopic quantum entities and the corresponding macroscopic systems in charge of their measurement, and, on the other hand, the conceptual entities of our human language and the cognitive systems in charge of interacting with them, through a dynamics based on \emph{meaning}. As we have mentioned in Section~\ref{intro}, this analysis has been used to achieve pa particularly effective modeling of human cognition processes, including judgment and decision making, reasoning, memory and perception \citep{aertsaerts1995,aertsbroekaertsmets1999,gaboraaerts2002,aertsczachor2004,khrennikov1999,altmanspacher2002}. 

If this correspondence between quantum physical entities and conceptual entities is not merely accidental, but an expression of the nature of our physical reality, at a fundamental level, it is also possible that the behaviour of quantum physical entities can also be explained by assuming that their nature is not particle-like, or wave-like, or something intermediary between a particle and a wave, but \emph{conceptual}, in the sense that our physical world would ultimately, at a fundamental level, be governed by dynamics which depend on the meaning that the different physical entities would carry. According to the \emph{conceptuality interpretation} elaborated by one of us, the duality between bosonic and fermionic entities\footnote{Unlike bosons, which obey the Bose--Einstein statistic and are integer spin entities, fermions obey the Fermi--Dirac statistic and are fractional spin entities. In particular, they obey the Pauli exclusion principle, which gives stability to matter by preventing identical fermions from occupying the same quantum state.}
would then be, for example, a reflection of the duality between conceptual entities, which carry meaning, and cognitive entities, which are sensitive to meaning \cite{aerts2009b,aertsetal2020}.  
 
We cannot, of course, go into the merits of this perspective on reality, where human language and the corresponding cognitive processes are seen as just one of many cognitive-conceptual structures that would have emerged in the course of evolution, from the very first moments of existence of our universe \citep{aertssozzo2015,aertssassoli2018,aertssassoli2022}.\footnote{The conceptuality interpretation did not arise from a fundamental philosophical principle, as is often the case with other interpretations, but from the analysis of concrete paradoxical quantum situations and the attempts to make sense of them. Fundamental to the interpretation is the insight that if bosons, the force-carrying quantum particles, are not objects but conceptual entities, then these paradoxical situations suddenly become intelligible. In this sense, the conceptuality interpretation is a bottom-up interpretation that proposes a radically new ontology for quantum entities. This makes it difficult to compare with the other existing quantum interpretations, none of which, in our view, provides a good explanation for all known quantum phenomena.}

We do not know whether such a perspective will find some compelling experimental confirmation in the future. At present, it remains a speculative hypothesis that has the advantage of offering numerous powerful explanations for the otherwise incomprehensible behavior of microscopic quantum systems. Here, we will use it to help us reason about the separability issue, aware that it may turn out to be much more than just a suggestive analogy, in the future.

So, according to the conceptuality interpretation, we will assume that a microscopic entity is conceptual in nature. Of course, by this we mean an entity that possesses the same nature as a human concept, even though it is not a human concept. This is a confusion we must not make: human concepts and microscopic entities are entirely different, just as a sound wave is completely different from an electromagnetic wave, even though they share the same undulatory nature. So, what we will do, in what follows, is to reason about human concepts assuming that what we can say about them can somehow be transferred  to the quantum microscopic entities, by virtue of the fact that they too would possess a conceptual-like nature. 

Let us assume that human culture, at some point in its history, was entirely transferred into an immense corpus of texts accessible online. These texts are to be understood as true \emph{coherence domains}, in the sense that each text in the corpus possesses its own cohesion, with all its parts connected to each other in terms of meaning, and it is precisely because of this internal cohesion through meaning that a text is distinguished from other texts in the collection, which express different coherence domains, i.e., different \emph{meaning domains}. We can think of this corpus as an evolution of the current World Wide Web \citep{qweb2018}, and we can imagine that we are able to interrogate it in order to gather knowledge about the conceptual entities of our language, whose semantic structure results from the combined cognitive activity of a large number of human beings. For instance, we can do so by assessing the number of texts jointly containing certain terms, and by doing so evaluate the strength of the  \emph{meaning connection} existing between them \citep{acds2010,a2011,aabbssv2017}. 

Note that when two conceptual entities are connected through meaning, it is in general possible to devise experiments through which the CHSH-Bell's inequalities \citep{chsh1969} can be violated \citep{as2011,bg2019,aabgssv2019}. When described via the quantum formalism, this means that these conceptual entities are typically in a state of entanglement, and following the analysis presented in the previous sections, we know they cannot be considered separate entities. In other words, when we are in the presence of a meaning bond \citep{a2011} that leads certain words, or word combinations, to cluster more easily in some texts, the conceptual entities associated with these words, or word combinations, cannot be considered separate. 

We will not go into detail about how meaning connections between certain concepts, i.e., between the words that represent them in our language, can be exploited to violate CHSH-Bell's inequalities, in well-conceived experiments. What is important for our discussion is the correspondence between the notion of \emph{quantum entanglement} and the notion of \emph{meaning connection}, with the latter appearing, for example, through a more frequent clustering of certain words, or word combinations, in the texts of our idealized documental entity, representing the totality of human written knowledge at a given moment of human evolution. This clustering effect, due to the existence of meaning connections, can also be understood as resulting from a mechanism of \emph{contextual updating}. In fact, whenever, in writing a text, we add a new word, it cannot be entirely arbitrary, as it will depend on the meaning of all the words that have already been written. And through the addition of a new word, the meaning of the entire text will also in turn be contextually updated. This mechanism, precisely because it works on the meaning connections that exist between words, and word combinations, is an expression of the lack of independence of the latter, which tend to cluster together and, in doing so, reveal the conceptual-cognitive equivalent of the behavior of a Bose--Einstein gas \citep{ijtp2023,philtransa2023,aertsetal2024paper1}.   

If we consider the above, we are tempted to say that there is no separation between the different conceptual entities belonging to our human language, and if our conceptualist analogy is valid, neither can there be true separation between physical entities, since every physical entity would be in essence quantum, and the non-classical behavior of a quantum entity would be the result of its conceptual nature. However, things are not so simple. In fact, it is one thing to consider conceptual entities described by single words, or combinations of a few words, as may be the case with a short sentence, and quite another to consider more extended linguistic entities, such as entire paragraphs, pages, or even whole stories. In our conceptualist analogy, these entities, composed of an increasing number of single-word conceptual entities, are the equivalent of aggregates of  interconnected microscopic entities forming a coherent and cohesive whole that we usually call a macroscopic body, behaving according to the laws of classical mechanics. 

Historically, we have always thought of spatially separate macroscopic bodies as separate systems, and that is why Einstein, and many of his colleagues of his time, were not willing to accept quantum non-separability so easily. In their experience, both as human beings and as physicists, they had always observed that physical objects are perfect examples of separate entities and that even our theoretical constructions, until the advent of the quantum revolution, had always been based on the independence of systems, when not connected by specific intermediary structures and separate by sufficiently large spatial distances (see our discussion in Section~\ref{intro}). But the separation of macroscopic bodies may be only apparent, in the sense that they may remain connected in a way that we are simply unable to detect, for example because, according to \emph{decoherence theory} \citep{joos1985}, entanglement would not disappear but only be ``diluted'' into the environment.\footnote{Decoherence is a consequence of the random photonic bombardment that exists in the hot environment in which we live, not only on this planet, due to the proximity of the Sun, but also in outer space, because of the cosmic background radiation left over from the Big Bang (about 2.76 kelvins).} Thus, non-separability would remain, simply becoming weaker because the entanglement connections would be distributed over all degrees of freedom in the environment. 

But let us push our conceptualist analogy further. The texts that form our idealized corpus of documents, which correspond to all the stories humanity has told up to the historical moment considered, constitute the semantic space in which we can test the presence of meaning connections between different concepts, by observing their joint presence and their greater or lesser proximity within these different texts, or stories (we use these two terms interchangeably). It is important to note that a text determines a state for the conceptual entities it contains, and the same is true for each part of a text: chapters, sections, pages, paragraphs, sentences, word combinations, they are all to be considered possible states for the conceptual entities (e.g., the single words) they contain \citep{aertssassolisozzo2016}. In other words, the presence of two concepts within a same text is already revealing of the fact that that a meaning  connection exists between them, whose strength will depend on their distance in the text and on the frequency of their appearance in the whole of the texts forming the documental entity.\footnote{The joint presence of terms representing specific concepts in a very large number of texts of the documental entity can be seen as an expression of a process of dilution of entanglement in the environment, due to the aforementioned phenomenon of quantum decoherence, to which corresponds a weakening of the meaning connection between the entities in question.}

Now, when the number of conceptual combinations increases, that is, when two conceptual entities are described by combinations of numerous words, or even phrases, paragraphs, etc., there is no guarantee that they will be found, jointly, within any of the texts of our documental entity, no matter how immense it is. And the greater the number of combinations, that is, the number of words that make up the two conceptual entities, the more likely it will be that they will be jointly found in no document. In the idealized situation we are considering, where human culture has been entirely transferred into such a huge body of texts, we can imagine that each person who writes, does so by directly transferring his or her text into the documental entity. And to make our example even more pregnant, we can imagine that even conversations and thoughts are directly transferred into the latter, in real time.\footnote{This example allows us to better express our point. This is not to say that the authors wish to see a society where such a possibility will be implemented, with all the privacy problems that this entails. Unfortunately, in the age of neurotechnology in which we live, such a scenario is far from hypothetical \citep{farahany2023}.} Thus, every linguistic creation by an intelligent cognitive entity, human or artificial, will be found in the corpus as a text concretizing a specific coherence-meaning domain. On the other hand, however immense this collection of texts is, their number still remains finite, thus their content limited, and it is easy to imagine that it will not contain every possible combination of stories that have been written, told or thought. And the longer these stories are, the more likely it is that they will not appear, jointly, in a single text of the corpus, hence, no meaning connections will exist between them and we can then assert that the conceptual entities associated with these stories are genuinely separate. 

If we take our conceptualist analogy seriously, we can thus say that just as certain stories, or fragments of stories, do not appear jointly in individual texts of our documental entity, as such texts have not yet been written, for example because they would hardly be regarded as significant if the stories in question possess meanings that are too distant,  \emph{mutatis mutandis} we can assume that there are composite physical entities whose size of the sub-entities forming them is such that there are no entangled states for them, thus, they are composite entities formed by genuinely separate sub-entities, impossible to describe within the standard quantum formalism. And just as our conceptualist analogy allows us to say that if two human conceptual entities are separate, at a given epoch in the evolution of our human culture, this does not mean that they will remain so in the future, since as our culture evolves it may create new connections of meaning that did not exist before, giving birth to new texts that can contain them jointly, in the same coherence domain, the same would apply, \emph{mutatis mutandis}, to physical entities. Indeed, as we explained in Section~\ref{ss}, the notion of separation between two physical systems is not something that can be expressed regardless of the measurements that are available. It is the latter, in fact, that determine whether two entities are separate or not, and it is sufficient that there exists a single non-separate measurement to decree that two entities, or systems, are non-separate.  

Today we know how to put in entangled states not only elementary entities, such as photons and electrons, but also whole atoms \citep{borselli2021}, and more generally we have put in superposition states molecular systems consisting of hundreds of atoms \citep{gerlich2011}, and even small mechanical resonators \citep{oconnell2010}. To achieve these exploits one must, of course, be able, among other things, to bring systems to extremely low temperatures in nearly perfect vacuum situations. These extremely low temperatures are so close to the absolute zero that nowhere in the universe similar conditions are known to exist.\footnote{The coldest place in the known universe is believed to be, according to NASA, the Boomerang Nebula, with a temperature of one degree Kelvin; see {\it https://science.nasa.gov/missions/hubble/boomerang-nebula}.} And even lower temperatures are needed to obtain the collective coherent states describing Bose--Einstein condensates \citep{anderson1995}.  This suggests that the existence of humanity, and of similar intelligent species in the universe, is capable of shifting the ``cut'' that determines which entities are to be regarded as separate, and therefore classical, and which are instead to be regarded as non-separate, and therefore quantum, or quantum-like. 

The fact that it is possible to devise experiments that can put systems as those mentioned above, formed by aggregates of numerous sub-systems, in superposition/entangled states, does not mean that the limit is only technological. Indeed, it is not possible to know whether it is possible to prepare certain entities in entangled states until one is able to also put in place experiments capable of testing the correlations that such states can create. And it may be the case that, beyond a certain size of the clusters, such experiments are not anymore possible, given for instance the size of our universe. And if the experiments are not available, not even in principle, we don't have the ability to say that the corresponding entities are non-separate. 

One aspect that we have not analyzed in detail and mention here only briefly, as it complements our previous analysis, is that of communication between two cognitive entities. If we walk away from a group of friends who are chatting, there comes a time when we no longer understand what they are saying to each other. This happens even when we keep hearing the sound of their voices. In other words, the coherence of their conversation no longer reaches us, while the sound of their conversation continues to do so. 

We can easily understand why this is so. In order to grasp the coherence of the conversation between the group of friends, we have to be actively listening, filling those gaps corresponding to the meanings we cannot grasp, making use of the meanings we are able to grasp instead. However, this only works if the latter are sufficient in number because, otherwise, the whole understanding, thus the whole coherence, suddenly collapses. 

It is clear from this example how the coherence of understanding a story is constructed by both entities, the one that hears it and the one that produces it. According to this view, which we analyzed in more detail in \citet{aerts2014}, cognitively separated systems would be such because they no longer understand each other, even though they can still hear each other. Translated into the physical domain, these would be systems that, while still interacting, can no longer maintain quantum coherence, because there would be too many domains of coherence in one system that, by not reaching the other, prevent it from grasping the overall coherence, and vice versa. 

Summing up our considerations, if we believe that the standard interpretation of quantum mechanics provides the only admissible state space, which is the linear Hilbert space, then separate systems do not exist and in principle all physical entities can be brought into entangled states. For bodies formed by large aggregates of sub-entities, like say two chairs, connecting them through entanglement may be of course extremely difficult to do, but it would be in principle a possibility. In our conceptualist analogy, this means that we can always find a way to create a story, which conveys a well-defined meaning to its listeners, while containing in its narrative arbitrarily long and arbitrarily different stories. On the other hand, if we do not attach a priori to the linear Hilbert space structure a fundamental role, not believing, for example, that axiom 5 has universal validity, we can adopt the more cautious stance that considers that the issue cannot be decided a priori, hence the possibility of having genuine separate systems should remain open. Certainly, to investigate fundamental questions of this kind, it is preferable not to limit the structure of the space of states from the outset, hence to consider that more general mathematical structures can play a role in the investigation of the different systems that belong to our reality.

Returning one last time to our conceptualist analogy, we can ask:  Is the ``Gulliver's travels'' story, written by Jonathan Swift, separate from the ``Winnie the Pooh'' story, written by Alan Alexander Milne?  One would be tempted to say yes, because the two stories, when brought artificially together, will not be coherent in the same way they are coherent when considered separately. Also, it is extremely difficult to imagine a longer narrative containing them verbatim and still preserving the coherence of its meaning domain. Thus, our conceptualist analogy would lead us to say that texts, beyond a certain number of words, naturally separate from the point of view of their meanings, hence the same should apply to material entities. But of course, it is also possible to object that, at the level of human culture, everything is always in principle connectable, because all stories arise from human writers who learned to speak a language formed by symbols, etc., and this very article, which brings the titles of these two stories together, can be seen as the embryo of a narrative that in the future could encapsulate them into a perfectly coherent whole. 

This new narrative, in which two initially disconnected stories can give rise to a new and larger domain of coherence, is an expression of the human capacity to interweave new structures of meaning, connecting what was hitherto disconnected. Still in the context of human culture, however, we can observe that the opposite phenomenon is also present, namely the destruction of pre-existing meaning structures. As we explained in \citet{aertssozzo2015,aertssassoli2022}, the destruction of knowledge through the destruction of meaning connections is the cognitive equivalent of the destruction of coherence due to heat in the physical domain, that is, the presence of the ubiquitous heat photons that bombard every structure. Despite this, biological life first, and human culture later, have managed to partially protect themselves from this incessant heat bombardment, restoring in some specific domains the presence of high concentrations of coherence and meaning. 

Just as it is possible to protect certain domains, others, those unable to fight against the growth of entropy, i.e., against the action of the second law of thermodynamics, will inexorably be driven towards stable states of equilibrium. In our human cognitive domain, we can consider the example of a city, where we find both high concentrations of meaning, for example in the city's library, but also areas where such meaning has been destroyed, for example where city's dumps are located, which are places where cultural artifacts are dismembered, losing once and for all their original coherence. In fact, it is no longer possible to reconstruct the story told in a book from only a few torn pages of it. Hence, if these pages lie far apart in the narrative, they will de facto become two separate entities. 

Most of our macroscopic universe can be likened to city's dumps. In fact, when the random photon bombardment is considered, it is no longer even a matter of distinguishing between macroscopic and microscopic systems, as the latter too are deeply affected by a noisy environment. Once again, we can ask whether decorrelating agents in the environment, like those in a thermal bath, are able to produce a genuine separation between two systems, be them macroscopic or microscopic, or whether entangled states persist, but in a diluted form, hence without our human measuring instruments being able to highlight them. In other words, does entanglement just weaken, in the sense that at all times, a portion of it, however small, still remains, or is it suppressed at some point altogether? If we consider the phenomenon of {\it entanglement sudden death}, also called {\it early-stage disentanglement}, i.e., the fact that even weakly dissipative environment are able to degrade quantum correlation to zero in finite time \citep{Yueberly2009,Bakshi2024}, rather than according to an exponential decay law, as is the case for classical correlations, the answer seems to be that a noisy environment is actually able to fully suppress entanglement between two or multiple systems, after a finite amount of time, hence bringing them to effective separation.\footnote{The situation is not so simple when it comes to describing more than two qubits, as there is no finite algorithm available to decide whether a given state, described by a density operator, is entangled or not \citep{mintert2005}.}

\section{Conclusion\label{con}}

Henri Poincar\'e famously wrote \citep{poincare1902}: ``If the different parts of the universe were not like the members of one body, they would not act on one another, they would know nothing of one another; and we in particular would know only one of these parts. We do not ask, then, if nature is one, but how it is one.''

In this article, building on Howard's historical analysis  of Einstein's concerns related not so much to Heisenberg's uncertainty principle, but to the lack of separation between quantum systems, we asked about the nature and structure of the ``one'' that Poincar\'e spoke of. Is it an entirely holistic-interconnected reality, where there is no possibility of separation between the interacting parts, or is it a mixture of both connected and separate parts? 

In answering this question, we have highlighted what the structural limitation of standard quantum mechanics is, when it is described using an axiomatic approach that does not take the Hilbertian structure for granted. Two of its axioms (covering law and weak modularity) in fact prevent the description of separate systems, in the sense that they are manifestly violated by them. We have also shown that the problem of the inability of the quantum formalism to describe an experimental separation between two systems does not appear at the level of states, which are in overabundance, but at the level of properties, and measurements, which are instead in insufficient number. In a sense, the superabundance of states, due to the superposition principle, generates the shortage of properties, described in the standard quantum formalism by orthogonal projectors. 

In our analysis, we also pointed out that the EPR article, and similar reasonings proposed by Einstein, pose a logical problem, which remains so regardless of the outcomes of experiments like those testing the CHSH-Bell inequalities. If we believe that a physical theory must be able to also describe separate systems, then quantum mechanics is necessarily incomplete. Indeed, it fails to describe situations like two spatially separate macroscopic bodies at room temperature, whose measurements are evidently separate. It fails to do so in the same way it fails to demarcate, at a fundamental level, the difference between a measuring instrument and the measured entity ({\it Heisenberg's cut}).

Einstein was however wrong in his belief that spatial separation would necessarily imply, for all physical entities, a separation of their measurements. There are certainly entities that can be prepared in states such that they are not anymore separate, even though there is a large spatial distance between them. All effects aren't local, and joint measurements on composite entities can be non-separate measurements, even if they are light years apart, as countless experiments have now demonstrated. But Einstein was right in the general idea that a non-separable quantum mechanics is necessarily incomplete when applied to systems assumed to be separable, although he did not have the mathematical and conceptual tools to prove such a claim in all rigor and generality. 

The question remains as to whether entangled (superposition) states, which are at the origin of quantum correlations and prevent the existence of separate measurements, are ubiquitous or not. Is it always possible, in principle, to put any pair of physical systems in an entangled state, or there are  instead  situations where two systems, such as planet Earth and our Sun, can be said to be authentically separate? In order to shed some light on this difficult question, we proposed an analogy with the semantic space formed by all texts produced by our human activity, assuming that quantum entities and conceptual entities would share, metaphysically speaking, a same nature, as suggested by the conceptuality interpretation of quantum mechanics. 

Our conceptualist analogy allowed us to observe that it is not possible to demarcate the two possibilities. Perhaps our physical reality possesses a {\it perfectly holistic structure}, where everything is interconnected (non-separate) to everything else, or perhaps things are more complex and at any given historical moment there are both separate and non-separate entities, with the demarcation between separation and non-separation being able to evolve over time. In other words, in considering issues of this kind, the suggestion is not to take a priori fideistic positions and to work, where possible, with formalisms capable of describing, at an ontological level, both separation and non-separation. This is also strongly suggested by the analysis of systems subject to environmental noise, where entanglement appears to be able to fully degrade in finite times. Hence, a description that takes into account a blending of these two structural possibilities is to be preferred, also because it could be the one that, in the future, will allow us to solve as yet unsolved foundational problems, such as the failure to elaborate a quantum field theory of gravity in a way that makes general relativity theory and quantum mechanics compatible.

\section*{Acknowledgements}
This work was supported by the project ``New Methodologies for Information Access and Retrieval with Applications to the Digital Humanities'', scientist in charge S. Sozzo, financed within the fund ``DIUM -- Department of Excellence 2023--27'' and by the funds that remained at the Vrije Universiteit Brussel at the completion of the ``QUARTZ (Quantum Information Access and Retrieval Theory)'' project, part of the ``Marie Sklodowska-Curie Innovative Training Network 721321'' of the ``European Unions Horizon 2020'' research and innovation program, with Diederik Aerts as principle investigator for the Brussels part of the network.

\end{document}